\documentclass[prd,aps,preprint,showpacs,nofootinbib,eqsecnum,superscriptaddress]{revtex4-1}
\usepackage[utf8]{inputenc}
\usepackage{hyperref}
\usepackage{amsfonts,amssymb,amsmath,amsthm,amsxtra,mathtools,euscript,mathrsfs}

\usepackage{bm}
\usepackage{graphicx,color,xcolor,subfigure}
\usepackage{dcolumn}
\usepackage{bbm}
\usepackage{multirow}

\begin{document}
\title{Cosmological Constraints on Non-flat Exponential $f(R)$ Gravity}
	
\author{Chao-Qiang Geng}
\email[Electronic address: ]{geng@phys.nthu.edu.tw}
\affiliation{School of Fundamental Physics and Mathematical Sciences, Hangzhou Institute for Advanced Study, UCAS, Hangzhou 310024, China}
\affiliation{International Centre for Theoretical Physics Asia-Pacific, Beijing/Hangzhou, China}
\affiliation{Department of Physics, National Tsing Hua University, Hsinchu 300, Taiwan}
\author{Yan-Ting Hsu}
\email[Electronic address: ]{ythsu@gapp.nthu.edu.tw}
\affiliation{Department of Physics,
	National Tsing Hua University, Hsinchu 300, Taiwan}
\author{Jhih-Rong Lu}
\email[Electronic address: ]{jhih-ronglu@gapp.nthu.edu.tw}
\affiliation{Department of Physics, National Tsing Hua University, Hsinchu 300, Taiwan}	

\begin{abstract}
We explore the viable $f(R)$ gravity models in FLRW backgrounds with a free spatial curvature  parameter $\Omega_{K}$.
In our numerical calculation, we concentrate on the exponential  $f(R)$ model of $f(R) = R - \lambda R_{ch}(1-\exp{(-R/R_{ch}}))$, where  $R_{ch}$ 
is  the characteristic curvature  scale, which is independent of  $\Omega_K$, and $\lambda$ corresponds to  the   model parameter,  while  $R_{ch}\lambda=2\Lambda$ with $\Lambda$ the cosmological constant.
  In particular, we study the evolutions of the dark energy density  and equation of state for  exponential $f(R)$ gravity 
  in  open, flat and closed universe, and compare with those for $\Lambda$CDM. From the current observational data, we find that  
   $\lambda^{-1}=0.42927^{+0.39921}_{-0.32927}$ at 68$\%$ C.L and $\Omega_K=-0.00050^{+0.00420}_{-0.00414}$ at 95$\%$ C.L.
  in the exponential $f(R)$ model.
  By using Akaike information criterion (AIC), Bayesian information criterion (BIC) and Deviance Information Criterion (DIC), 
  we conclude that there is no strong preference between the exponential $f(R)$ gravity and $\Lambda$CDM models  in the non-flat universe.

\end{abstract}
\maketitle
	
\section{Introduction}
Cosmological observations hint that our universe has been experiencing another accelerating expansion in the recent epoch~\cite{SupernovaSearchTeam:1998fmf,SupernovaCosmologyProject:1998vns} besides inflation in the very early time. 
However,   the origin of the late time acceleration remains a mystery. Although the $\Lambda$CDM model, in which the  cosmological constant $\Lambda$ plays the role of  dark energy, could give an explanation about this problem, it still suffers from some difficulties, such as
the cosmological constant problem~\cite{Weinberg:1988cp,Peebles:2002gy} and Hubble tension~\cite{Riess:2019cxk}. 
To describe our accelerating universe, many models with dynamical dark energy~\cite{Copeland:2006wr} 
beyond $\Lambda$CDM have been proposed.
In particular, there are two representative  approaches, in which one is to introduce some unknown matters called ``dark energy'' 
in the framework of general relativity~\cite{Copeland:2006wr, Li:2011sd}), and the other 
 is to modify the gravitational theory, e.g., $f(R)$ gravity~\cite{Nojiri:2010wj, Sotiriou:2008rp,DeFelice:2010aj}. 


It is known that  $f(R)$ gravity replaces the Ricci scalar, $R$, in the Einstein-Hilbert action with an arbitrary function of $f(R)$. Several viable models have been constructed in $f(R)$ gravity~\cite{DeFelice:2010aj,Bamba:2010iy}, such as Starobinsky~\cite{Starobinsky:2007hu}, Hu-Sawiki~\cite{Hu:2007nk}, Tsujikawa~\cite{Tsujikawa:2007xu,Cen:2019ohm}, and exponential~\cite{Linder:2009jz,Bamba:2010ws}  models.
These models  satisfy the following viable conditions~\cite{DeFelice:2010aj,Bamba:2010iy}: (1) the positivity of effective gravitational couplings;
(2) the stability of cosmological perturbations; (3) the asymptotic behavior to $\Lambda\mathrm{CDM}$ in the large curvature regime;
(4) the stability of the late-time de Sitter point; (5) constraints from the equivalence principle;
and (6) solar-system constraints. 
In this study, to illustrate our numerical results we concentrate on the exponential $f(R)$ gravity model, 
which contains only one more parameter than the standard $\Lambda$CDM model of cosmology.

Recently,  the  survey of the Planck 2018 CMB data along with  $\Lambda$CDM has
 suggested that our universe is closed at 99\% C.L.~\cite{DiValentino:2019qzk}. Motivated by this result, we would like to examine 
the  viable $f(R)$ gravity models without the spatial flatness assumption and explore  the constraints on  the models from the recent observational data. 
We  would also compare  viable $f(R)$ gravity with $\Lambda$CDM with the
spatial curvature  parameter
$\Omega_{K}$ set to be  free. We note that the study of the  viable $f(R)$ gravity models with an arbitrary  spatial curvature has not been performed in the literature yet.  
To illustrate our results, we will concentrate on the viable exponential $f(R)$ model.
	
The paper is organized as follows. In Section~\uppercase\expandafter{\romannumeral 2}, we review the Friedmann equations in $f(R)$ gravity in the non-flat background.  In Section~\uppercase\expandafter{\romannumeral 3}, we present the cosmological evolutions of the dark energy density 
parameter and equation of state in open, flat and closed exponential $f(R)$
gravity models and constrain the model parameters by using 
the Markov Chain Monte Carlo (MCMC) method. We summarize our results in Section~\uppercase\expandafter{\romannumeral 4}.

\section{$f(R)$ gravity  in spatially Non-Flat FLRW Spacetime} 
	
The action of  $f(R)$ gravity is given by
	
\begin{align} \label{frAction}
S=\int d^4 x \frac{\sqrt{-g}}{2\kappa^2}f(R) +S_{M},
\end{align}
where $\kappa^2 = 8 \pi G$ with $G$ the Newton's constant, $S_{M}$ is the action for both relativistic and non-relativistic matter. 
In the viable exponential gravity model, $f(R)$ is given by~\cite{Zhang:2005vt,Tsujikawa:2007xu,Linder:2009jz,Bamba:2010ws,Yang:2010xq}
\begin{align}
\label{Exp}
f(R) = R - \lambda R_{ch}(1-e^{-R/R_{ch}})
\end{align}
where $R_{ch}$ is related to the characteristic curvature modification scale.
Based on  the viable $f(R)$ conditions, one has that, when $R \rightarrow \infty, \, f(R) \rightarrow R-2\Lambda$, the product of $\lambda$ and $R_{ch}$ corresponds to the cosmological constant by the relation of  $\lambda R_{ch} = 2\Lambda$. As a result,
there is only one additional model parameter in the exponential gravity model in \eqref{Exp}.

Varying the action \eqref{frAction}, field equations for $f(R)$ gravity can be found to be

\begin{align} \label{eom}
FR_{\mu\nu}-\frac{1}{2}g_{\mu\nu}f -\nabla_{\mu}\nabla_{\nu}F+g_{\mu\nu}\square F=\kappa^2 T_{\mu\nu}^{(M)},
\end{align}
where $F\equiv df(R)/dR$, and $\square \equiv g^{\mu \nu}\nabla_{\mu}\nabla_{\nu}$ is the d'Alembert operator, and $T_{\mu\nu}^{(M)}$ represents the energy-momentum tensor for relativistic and non-relativistic matter. The above equation \eqref{eom} can also be written as 
\begin{align}
    G_{\mu\nu} = \kappa^2 \bigg(T_{\mu\nu}^{(M)} + T_{\mu\nu}^{(de)}\bigg)
\end{align}
where $G_{\mu\nu}=R_{\mu\nu}-(1/2)g_{\mu\nu}R$ is the Einstein tensor and
\begin{align}
    T_{\mu\nu}^{(de)} = \frac{1}{\kappa^2}\bigg(G_{\mu\nu}-FR_{\mu\nu} + \frac{1}{2} g_{\mu\nu} f + \nabla_{\mu}\nabla_{\nu}F - g_{\mu\nu}\square F\bigg)
\end{align}
stands for the energy-momentum tensor for dark energy.
\subsection{Modified Friedmann Equations}
We consider the spatially non-flat Friedmann-Lemaitre-Robertson-Walker (FLRW) spacetime, given by
\begin{equation}\label{flrw}
ds^2=-dt^2+a^{2}(t)\bigg(\frac{dr^2}{1-Kr^2}+r^2 d \theta^2+r^2 sin^2\theta d\phi^2\bigg),
\end{equation}
where $a(t)$ is the scale factor, and $K=-1,0,1$ represent the spatially open, flat and closed universe, respectively. 
Applying the metric \eqref{flrw} into \eqref{eom}, we obtain the modified Friedmann equations, given by
\begin{align}
&3FH^2+\frac{3KF}{a^2}=\frac{1}{2}(FR-f)-3H\dot{F}+\kappa^2\rho_M,\label{feg1}\\
&\ddot{F}=H\dot{F}-2F\dot{H}+\frac{2KF}{a^2}-\kappa^2 (\rho_M+P_M), \label{feg2}
\end{align}
where $H=\dot{a}/a$ is the Hubble parameter, the dot ``$\cdot$'' denotes the derivative w.r.t the cosmic time $t$, and the Ricci scalar $R$ takes the form
\begin{align}
R=12H^2+6\dot{H}+\frac{6K}{a^2}.
\end{align}
In order to study the behavior of dark energy and the effects of  spatial curvature, we rewrite the modified Friedmann equations in \eqref{feg1}
and  \eqref{feg2} as
\begin{align}
\label{H2}
H^2&=\frac{\kappa^2}{3}(\rho_M+\rho_{de}+\rho_K),\\  
\label{Hdot}
\dot{H}&=-\frac{\kappa^2}{2}(\rho_M+\rho_{de}+\rho_K+P_M+P_{de}+P_K),
\end{align}
where $\rho_M = \rho_m + \rho_r$ is the density of  non-relativistic matter and radiation, 
while the dark energy density and pressure are given by
\begin{align}
\label{eqrhode}
\rho_{de}&=\frac{3}{\kappa^2}\bigg(H^2(1-F)-\frac{1}{6}(f-FR)-H\dot{F}+\frac{K}{a^2}(1-F)\bigg),\\
P_{de}&=\frac{1}{\kappa^2}\bigg(\ddot{F}+2H\dot{F}+\frac{1}{2}(f-FR)-(1-F)\big(3H^2+2\dot{H}+\frac{K}{a^2}\big)\bigg),
\end{align}
respectively. Here, the effects of   spatial curvature in  the modified
Friedmann equations can be described by the effective energy density and pressure, written as
\begin{align}
\rho_K &=-\frac{3K}{\kappa^2 a^2}, \label{rhoK}\\
P_K &= \frac{K}{\kappa^2 a^2}
\end{align}
respectively.
Note that the energy density and pressure for  non-relativistic matter, radiation,  dark energy and   spatial curvature satisfy the continuity equation
\begin{align}
    \frac{d \rho_{i}}{dt} + 3H(1+w_{i})P_{i}=0
\end{align}
where   $w_{i}$ with  $i = (m,r,de, K)$ represent the corresponding equations of state, defined  by
\begin{align}
    w_{i} \equiv \frac{P_{i}}{\rho_{i}}\,,
\end{align}
 respectively.
By rewriting the Friedmann equation of \eqref{H2} in terms of observational parameters, we have 
\begin{align}
    1=\Omega_{m}+\Omega_{r}+\Omega_{de}+\Omega_{K}\,,
\end{align}
where $\Omega_{i}$ are the corresponding density parameters, defined by 
\begin{align}
\Omega_{i}=\frac{\kappa^2\rho_{i}}{3H^2}. 
\end{align}
From \eqref{rhoK}, we have that $\Omega_{K}=-K/(aH)^2$ with $\Omega_{K}>0 , =0$ and  $<0$ for open, flat and closed universe, respectively. 

In order to solve the modified Friedmann equations numerically, we define the dimensionless parameters $y_H$ and $y_R$ to be
\begin{align}
y_H &\equiv \frac{\rho_{de}}{\rho_m^{(0)}} = \frac{H^2}{m^2}-a^{-3}-\chi a^{-4} - \beta a^{-2},\\
y_R &= \frac{R}{m^2}-3a^{-3},
\end{align}
where $m^2=\kappa^2 \rho_m^{(0)}/3 $,  $\chi=\rho_r^{(0)}/\rho_m^{(0)}$, and $\beta=\rho_K^{(0)}/\rho_m^{(0)}$ with $\rho_i^{(0)}\equiv\rho_i(z=0)$.
It can be found from \eqref{feg1} that these two parameters obey the equations
\begin{align}
\frac{dy_H}{d \text{ln}a}&=\frac{y_R}{3}-4y_H,\\
\frac{dy_R}{d \text{ln}a}&=\frac{1}{m^2} \frac{dR}{d \text{ln}a}+9a^{-3}.
\end{align}
As a result, one is able to combine these two first order differential equations into a single second order equation
\begin{align}
\label{eq:ddyh}
y_H''+J_1y_H'+J_2 y_H +J_3 = 0,
\end{align} 
where the prime ``$\prime$'' denotes the derivative w.r.t $\text{ln} a$, and
\begin{align}
J_1 &=4+\frac{1}{y_H+a^{-3}+\chi a^{-4} +\beta a^{-2}}\frac{1-F}{6m^2 F,_R},\\
J_2 &=\frac{1}{y_H+a^{-3}+\chi a^{-4}+\beta a^{-2}}\frac{2-F}{3m^2 F,_R},\\
J_3 &=-3a^{-3} -\frac{(1-F)(a^{-3}+2\chi a^{-4})+(R-f)/3m^2}{y_H+a^{-3}+\chi a^{-4} +\beta a^{-2}}\frac{1}{6m^2 F,_R}.\label{eq:J3}
\end{align}
With the differential equation in  \eqref{eq:ddyh}, the cosmological evolution can be calculated through the various existing programs in the
literature.

\section{Numerical Calculations}
 In this section, we study  the background evolutions of the dark energy density parameter and equation of state for the exponential $f(R)$ model without the spatial flatness assumption. We  modify the {\bf{CAMB}}~\cite{Lewis:1999bs}
program at the background level and use the {\bf CosmoMC} \cite{Lewis:2002ah} package, which is a {\bf Markov Chain Monte Carlo (MCMC)} engine, to explore the cosmological parameter space and  constrain the exponential $f(R)$ model from the observational data.

\subsection{Cosmological evolution} 
\label{subsec:evol}

\begin{figure}[!htb]
	\includegraphics[width=1 \linewidth]{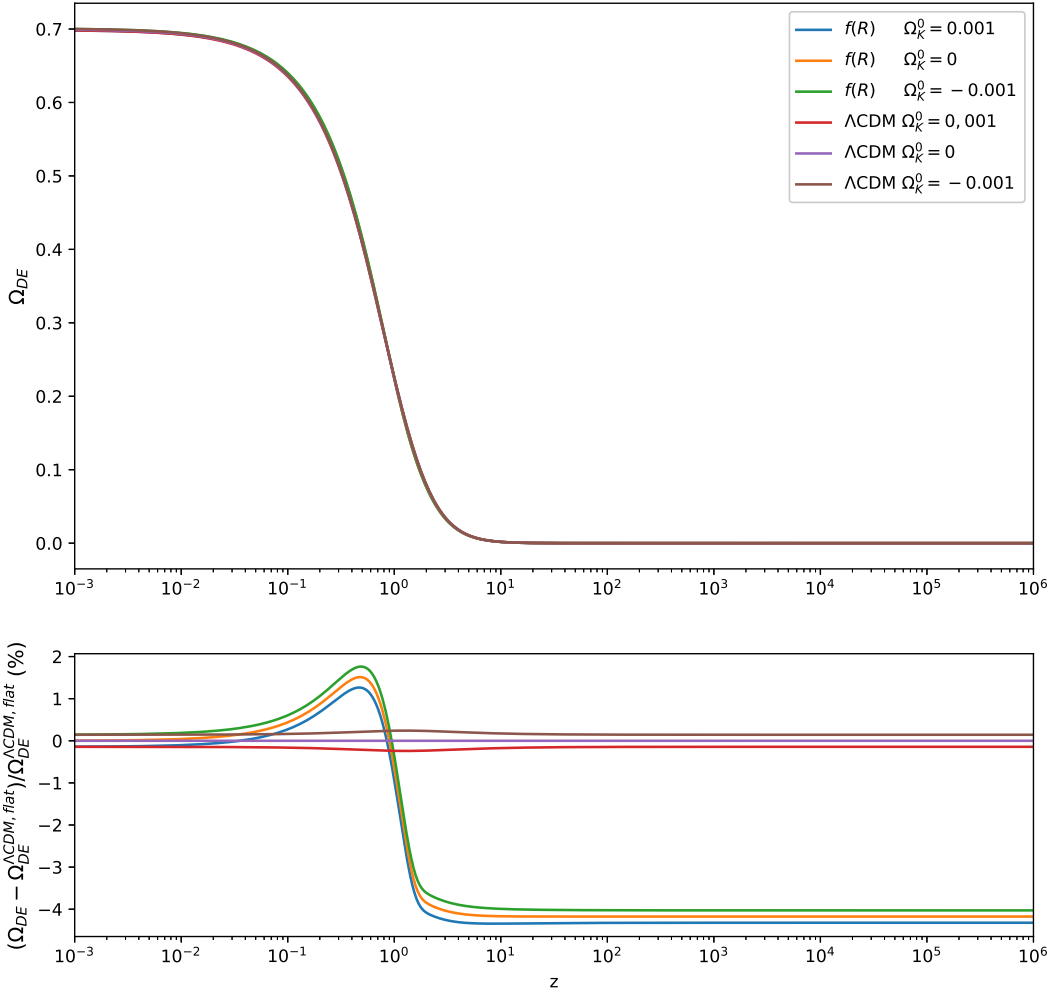}
	\caption{Evolutions of $\Omega_{DE}$ for exponential $f(R)$ gravity with $\lambda^{-1}=0.5$ and $\Lambda$CDM in open, flat and closed universe with $\Omega_{K}^0=(0.01,0,-0,01)$ (upper panel), and the residues with respect to flat $\Lambda$CDM
	$(\Omega_{DE}-\Omega_{DE}^{\Lambda CDM, flat})/\Omega_{DE}^{\Lambda CDM, flat}$
	(lower panel), where the initial values are give by  ($\Omega_{m}^0$, $\Omega_{r}^0$) = $(0.2998, 1.5 \times 10^{-3})$, $\Omega_{K}^0=(0.001,0,-0.001)$ and $H_0=67km/s/Mpc$. }
	\label{fig:omde}
\end{figure}

\begin{figure}[!htb]
	\includegraphics{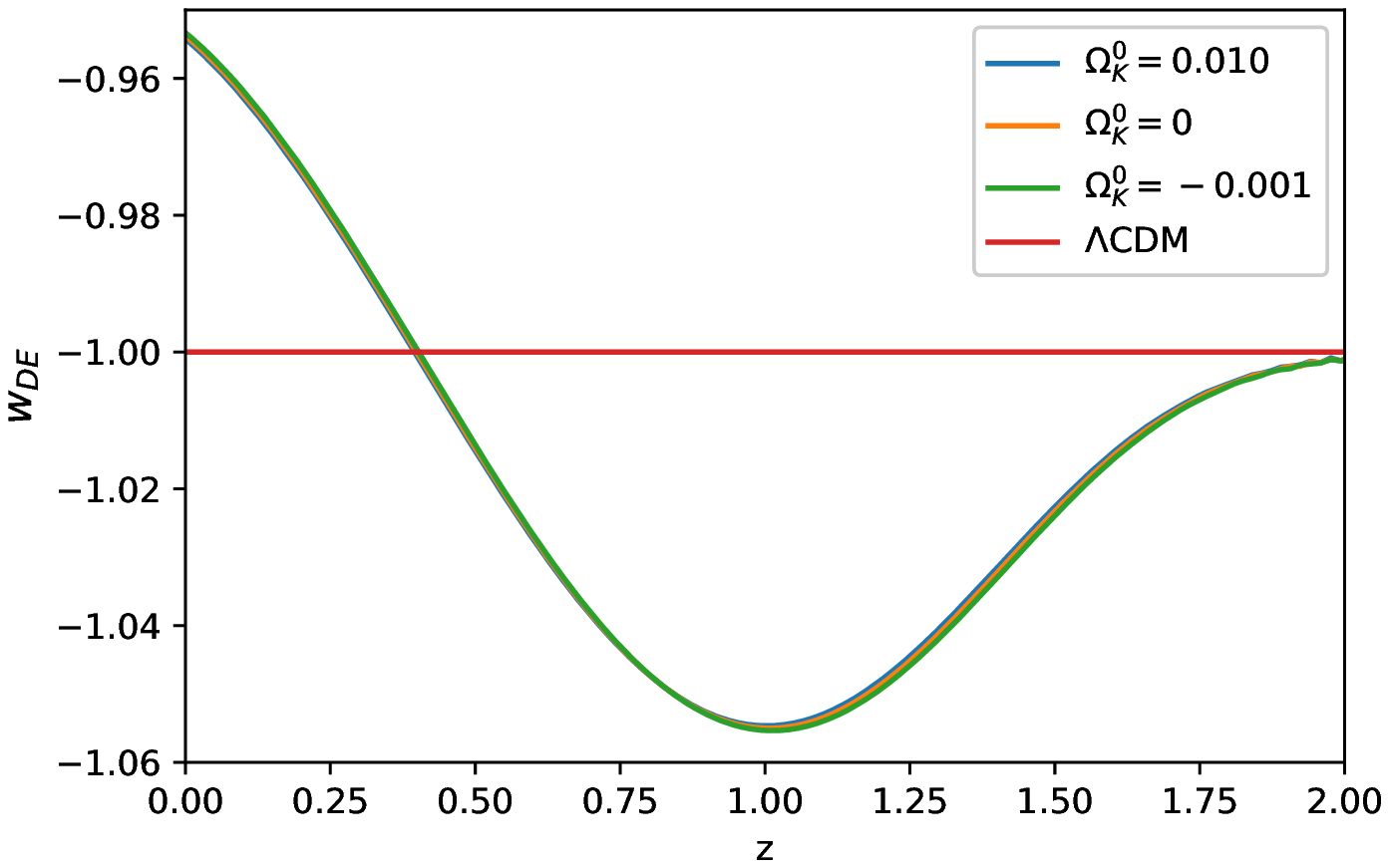}
	\caption{Evolutions of $w_{DE}$ for Exponential $f(R)$ gravity with $\lambda^{-1}=0.5$ and $\Omega_{K}^0=(0.01,0,-0,01)$ and $\Lambda$CDM,
	where the initial values are described in Fig.~\ref{fig:omde}. }
	\label{fig:wde}
\end{figure}

To examine the cosmological evolution of dark energy for the viable exponential gravity model, we plot 
the density parameter $\Omega_{DE}$, 
and equation of state $w_{DE}$ of the model in Figs.~\ref{fig:omde} and \ref{fig:wde}, respectively. 
From the previous studies of the exponential $f(R)$ gravity model, the model and spatial curvature density parameters are constrained to be  $0.392<\lambda^{-1}<0.851$ \cite{Chen:2019uci}
and  $-0.0011<\Omega_{K}^0<0.0027$ at 68$\%$ C.L.  \cite{Vagnozzi:2020rcz}, respectively. In this work, we choose $\lambda^{-1}=0.5$ and $\Omega_{K}^0=\pm 0.001$ to see the behavior of exponential $f(R)$ gravity.  The initial conditions are set to be ($\Omega_{m}^0$, $\Omega_{r}^0$) = $(0.2998, 1.5 \times 10^{-3})$, $\Omega_{K}^0=(0.001,0,-0.001)$ for the (open, flat, closed) universe, and $H_0=67$ km/s/Mpc.

Fig.~\ref{fig:omde} shows the evolutions of dark energy for exponential $f(R)$ gravity with $\lambda^{-1}=0.5$ and $\Lambda$CDM with $\Omega_{K}^0=(0.001,0,-0.001)$. From the figures, we see that the dark energy density parameter for exponential $f(R)$ gravity is slightly
larger (smaller) than that of $\Lambda$CDM when $10^{-1}\lesssim z \lesssim 10^{0}$ ($z \gtrsim 10^0$). 
It  approaches the cosmological constant in the high redshift region as a characteristic of the viable $f(R)$ gravity models. 
It is clear that the deviation for $f(R)$ gravity from flat $\Lambda$CDM is small, with $|(\Omega_{DE}-\Omega_{DE}^{\Lambda CDM, flat})/\Omega_{DE}^{\Lambda CDM, flat}| <  5 \%$. Note that $\Omega_{DE}$ for $f(R)$ gravity in the closed universe
has a larger value in comparison with the other cases. Fig.~\ref{fig:wde} illustrates equation of state $w_{DE}$ for dark energy as a function of $z$. We can see that $w_{DE}$ evolves from the phantom phase ($w_{DE}<-1$) to the non-phantom phase ($w_{DE}>-1$) for a fixed value of $\Omega_{K}^0$.


With the initial conditions and
\begin{align}
\label{eq:age}
t_{age}=\frac{1}{H_0}\int_{0}^{1} \frac{da}{a\sqrt{\Omega_{m}a^{-3}+\Omega_{r}a^{-4}+\Omega_{K}a^{-2}+\Omega_{de}(a)}}, 
\end{align} 
we can calculate the age of the universe. Consequently, we obtain that 
\begin{align}
\label{eq:age1}
t_{age}^{open}&=\quad14.021 \;,\  14.045 \quad Gyr \qquad   (\Omega_{K}^0=0.001)\\
\label{eq:age2}
t_{age}^{flat}&=\quad14.025    \;,\ 14.049 \quad Gyr \qquad (\Omega_{K}^0=0)\\
\label{eq:age3}
t_{age}^{closed}&=\quad14.028  \;,\ 14.054 \quad Gyr \qquad  (\Omega_{K}^0=-0.001),
\end{align}
for the exponential $f(R)$ and $\Lambda$CDM models in the open, flat and closed universe, respectively. 
From \eqref{eq:age1}, \eqref{eq:age2} and \eqref{eq:age3}, we see that
the age of the universe for exponential $f(R)$ gravity is shorter than the corresponding one for $\Lambda$CDM. 
Note that the bigger value of $t_{age}$ is related to the longer growth time of the large scale structure (LSS) and larger matter density fluctuations.


\subsection{Global fitting}
In this subsection, we present constraints on the cosmological parameters in the exponential $f(R)$ model  without the spatial flatness assumption . 
With the modifications of {\bf CAMB} at the background level and the {\bf CosmoMC} package, we perform the {\bf MCMC} analysis.

To break the geometrical degeneracy \cite{Efstathiou:1998xx,Howlett:2012mh}, we fit the model with the combinations of the observational data, including CMB temperature and polarization angular power spectra from 
{\it Planck} 2018 with high-$l$ TT, TE, EE , low-$l$ TT, EE, CMB lensing from SMICA~\cite{Planck:2018vyg,Planck:2018lbu,Planck:2019kim,Planck:2019nip}, BAO observations from 6-degree Field Galaxy Survey (6dF)~\cite{Beutler:2011gb}, 
SDSS DR7 Main Galaxy Sample (MGS)~\cite{Ross:2014qpa} and BOSS Data Release 12 (DR12)~\cite{BOSS:2016wmc}, 
and supernova (SN) data from the Pantheon compilation~\cite{Pan-STARRS1:2017jku}. As we set the density parameter of curvature and the neutrino mass sum to be free, our fitting for the exponential $f(R)$ model contains nine free parameters, where the priors are listed in Table~\ref{tab:prior}.

To find the best-fit results, we minimize the $\chi^2$ function, which is given by
\begin{eqnarray}
    	\chi^2 = \chi_{CMB}^2 + \chi_{BAO}^2
    	+ \chi_{Pan}^2\,.
\end{eqnarray}

Explicitly, we take
\begin{align}
    \chi_{CMB}^2 = \sum_{l l'}(C^{obs}_l - C^{th}_l) \mathcal{M}^{-1}_{l l'} (C^{obs}_{l'} - C^{th}_{l'})
\end{align}
where $C^{obs(th)}_l$ corresponds to the observational (theoretical) value of the related power spectrum,  and $\mathcal{M}$ is the covariance matrix for the CMB   data~\cite{Planck:2018vyg, Planck:2018lbu}.

For the BAO data, we adopt the dataset from 6-degree Field Galaxy Survey (6dF) at $z_{eff}= 0.106$~\cite{Beutler:2011gb}, 
SDSS DR7 Main Galaxy Sample (MGS) at $z_{eff}= 0.15$~\cite{Ross:2014qpa} and BOSS Data Release 12 (DR12) at $z_{eff}= (0.38,0.51,0.61)$ ~\cite{BOSS:2016wmc}. As a result, we have
\begin{align}\label{bao_chi2}
    \chi_{BAO}^2 = \chi_{6dF}^2 + \chi_{MGS}^2 + \chi_{DR12}^2.
\end{align}
For the uncorrelated data points, such as 6dF and MGS,  $\chi^2$ is given by
\begin{align}
    \chi^2(p) = \sum^N_{i=1} \frac{[A_{th}(z_i) - A_{obs}(z_i)]^2}{\sigma^2_i},
\end{align}
where $A_{th}(z_i)$ is the predicted value computed in the model under consideration, and $A_{obs}(z_i)$ denotes the measured value at  $z_i$ with
the standard deviation $\sigma_i$.
Note that in our study, we adopted 6dF and MGS, whose standard deviations are given by $\sigma_{6dF} = 0.015$ and $\sigma_{MGS} = 0.168$, respectively.
The data points from DR12 are correlated. In this case,  $\chi^2$ is given by
\begin{align}
    \chi_{DR12}^2 = \bigg[\textbf{A}_{th} - \textbf{A}_{obs}\bigg]^{T} \cdot \textbf{C}^{-1} \cdot \bigg[\textbf{A}_{th} - \textbf{A}_{obs}\bigg],
\end{align}
where $\textbf{C}^{-1}$ is the inverse of the covariance matrix, which is an $6\times 6$ matrix given in  Eq.~(20) of Ref.~\cite{Ryan:2019uor}.

For the Pantheon SN Ia samples, there are 1048 data points scattering  between $0.01 \leq z \leq 2.3$, with the observable to be the distance modulus $\mu$ defined in Ref.~\cite{Pan-STARRS1:2017jku}. We have that
\begin{align}
    \chi_{Pan}^2 = \bigg[\bm{\mu}_{obs} - \bm{\mu}_{th}\bigg]^{T} \cdot \textbf{Cov}^{-1} \cdot \bigg[\bm{\mu}_{obs} - \bm{\mu}_{th}\bigg],
\end{align}
where $\textbf{Cov}^{-1}$ is the inverse covariance matrix \cite{Pan-STARRS1:2017jku} of the sample including the contributions from both the statistical and systematic errors. The covariance matrix of Pantheon samples can also be found in the website \footnote{\url{http://supernova.lbl.gov/Union/}}.

\begin{table}[ht]
	\begin{center}
		\caption{This table contains priors of cosmological parameters for exponential
$f(R)$ and $\Lambda$CDM models. }
		\begin{tabular}{|c||c|} \hline
			Parameter & Prior
			\\ \hline
			$f(R)$ model parameter $\lambda^{-1}$& 
			$0.1 \leq \lambda^{-1} \leq 1$
			\\ \hline
			Curvature parameter $\Omega_K$& $-0.1 \leq \Omega_K \leq 0.1$
			\\ \hline
			Baryon density & $0.5 \leq 100\Omega_bh^2 \leq 10$
			\\ \hline
			CDM density & $0.1 \leq 100\Omega_ch^2 \leq 99$
			\\ \hline
			Optical depth & $0.01 \leq \tau \leq 0.8$
			\\ \hline
			Neutrino mass sum& $0 \leq \Sigma m_{\nu} \leq 2$~eV
			\\ \hline
			Angular size of the sound horizon  & $0.5 \leq 100 \theta_{MC} \leq 10$
			\\ \hline
			Scalar power spectrum amplitude & $1.61 \leq \ln \left( 10^{10} A_s \right) \leq 3.91$
			\\ \hline
			Spectral index & $0.8 \leq n_s \leq 1.2$
			\\ \hline
		\end{tabular}
		\label{tab:prior}
	\end{center}
\end{table}

\begin{table}[!htb] 
	\newcommand{\tabincell}[2]{\begin{tabular}{@{}#1@{}}#2\end{tabular}}	
	\begin{center}
		\caption{ The constraints of cosmological parameters for exponential 
$\Lambda$CDM models fitted with CMB+BAO+SN data sets, where the cosmological parameters are given at 95\% C.L, while $\Lambda^{-1}$ and $\Sigma m_{\nu}$ are given at 68\% C.L}	
		
		\scalebox{1.2}{
			\begin{tabular} {|c|c|c|}
				\hline
				Parameter&
				Exp $f(R)$&
				$\Lambda$CDM
				\\
				\hline
				
			    {\boldmath$\Omega_b h^2   $} & $0.02241^{+0.00032}_{-0.00031}$&
			    $0.02242^{+0.00031}_{-0.00031}$\\
			    
			    {\boldmath$\Omega_c h^2   $} & $0.11954^{+0.00270}_{-0.00266}$& $0.11948^{+0.00268}_{-0.00263}$\\

			    {\boldmath$\tau           $} & $0.05537^{+0.01518}_{-0.01424}   $& $0.05558^{+0.01567}_{-0.01407}   $\\
			    
			    {\boldmath$\Omega_K       $} & $0.00050^{+0.00420}_{-0.00414}$& $0.00050^{+0.00400}_{-0.00403}$\\
			    
			    {\boldmath$\Sigma m_\nu [eV]  $} & $< 0.06816                   $& $< 0.06121                  $\\
			    
			    {\boldmath$\lambda^{-1}   $} & $< 0.42927^{+0.39921}_{-0.32927}                   $&
			    $-$\\

			    {\boldmath$H_0 (km/s/Mpc)$}     &   $67.73344^{+1.39013}_{-1.45947}        $ & $67.95862^{+1.23489}_{-1.22664}        $\\

				${\rm{Age}}/{\rm{Gyr}}     $ & $13.7550^{+0.1585}_{-0.1581}     $ &
				$13.76^{+0.1578}_{-0.1514}     $\\
				\hline
				
				{\boldmath $\chi^2_{best-fit}$} & 
				$3821.50$& 
				$3821.84$\\
				
				\hline
		\end{tabular}}
		\label{tab:constraints}
	\end{center}
\end{table}
   
\begin{table}[ht]
	\begin{center}
	\caption{ The results of AIC, BIC and DIC computed from the sample we use for both $\Lambda$CDM and exponential $f(R)$ models, where 
	$\Delta AIC= AIC_{f(R)}-AIC_{\Lambda CDM}$, $\Delta BIC= BIC_{f(R)}-BIC_{\Lambda CDM}$, and $\Delta DIC= DIC_{f(R)}-DIC_{\Lambda CDM}$. }
		\begin{tabular}{|c||c|c|c|c|c|c|c|} \hline
			Model & $\chi^2_{min}$ & AIC & {$\Delta$AIC}  &  BIC & $\Delta$BIC &  DIC & $\Delta$DIC
			\\ \hline
			$\Lambda$CDM  & $3821.84$ & $3837.84$ & $0$ & $3887.35$ & $0$ & $3850.38$ & $0$ 
			\\ \hline
			Exp $f(R)$ & $3821.50$ & $3839.50$ & $1.66$ & $3895.20$ & $7.85$ & $3851.87$ & $1.49$ 
            \\ \hline
		\end{tabular}
		\label{tab:AIC}
	\end{center}
\end{table}   
   \begin{figure}[htb!]
   	\includegraphics[width=1 \linewidth]{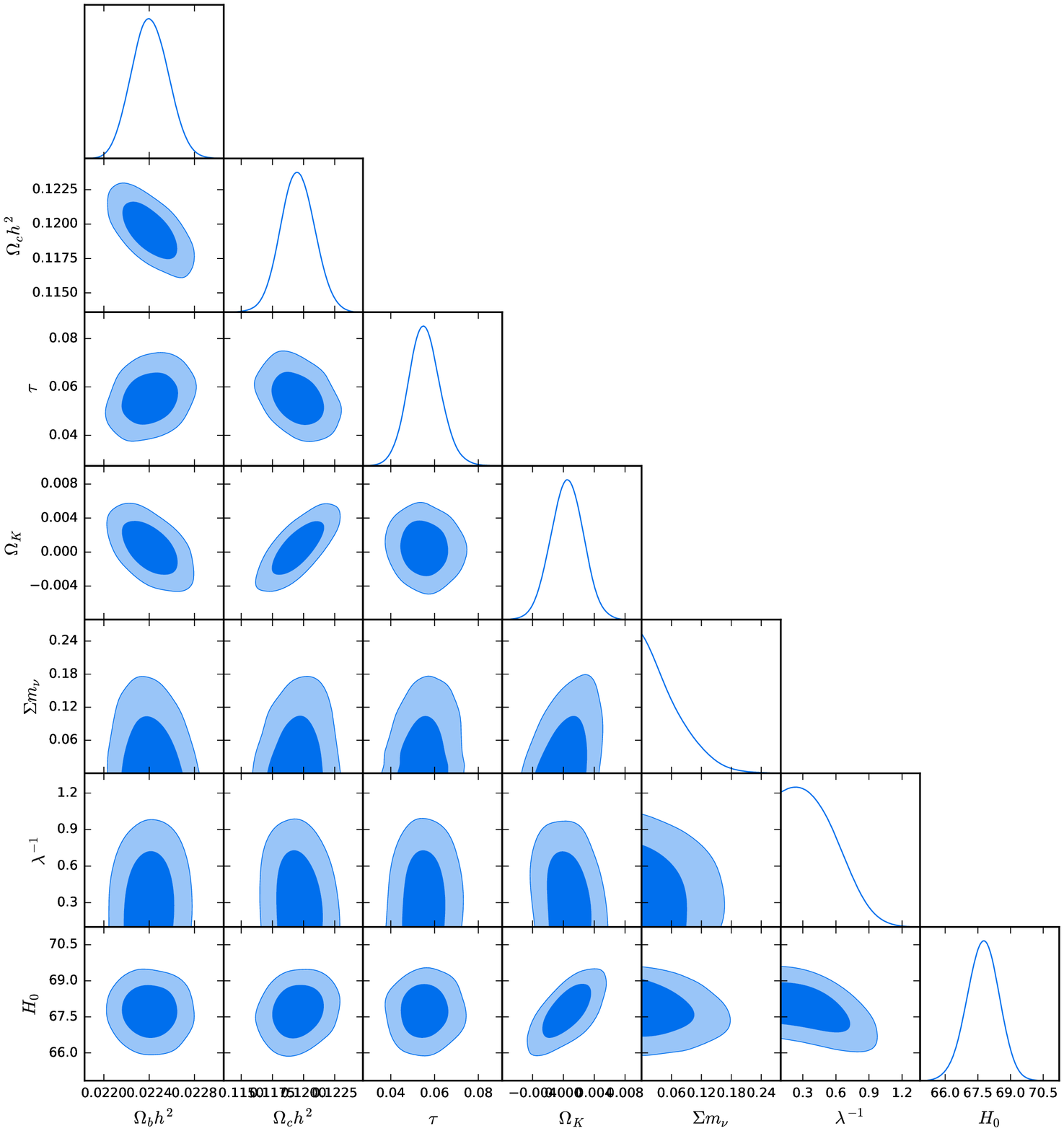}
   	\caption{Marginalized one and two-dimensional constraints of $\Omega_b h^2$, $\Omega_c h^2$, $\tau$, $\Omega_K$, $\sum m_\nu$, $\lambda^{-1}$ and $H_0$ for exponential $f(R)$ with the combined data of CMB+BAO+SN, where the contour lines represent 68$\%$~ and 95$\%$~ C.L., respectively. \label{fig:fit}}
   \end{figure}

The global fitting results of the exponential $f(R)$ model without the spatial flatness assumption with CMB+BAO+SN data sets are shown in Fig.~\ref{fig:fit} and listed in Table~\ref{tab:constraints}. We note that the exponential $f(R)$ model is barely distinguishable from  $\Lambda$CDM.
This statement is in agreement with the previous work in the flat universe~\cite{Geng:2014yoa}. 
FromTable~\ref{tab:constraints},  we see that  the model 
and curvature density parameters are constrained to be 
 $\lambda^{-1}=0.42927^{+0.39921}_{-0.32927}$ at 68$\%$ C.L and   $\Omega_K=-0.00050^{+0.00420}_{-0.00414}$ at 95$\%$ C.L for the exponential $f(R)$ model, respectively. Note that the flat $\Lambda$CDM model is recovered when $\lambda^{-1}=0$ and $\Omega_K=0$.
We also obtain that $\chi^2=3821.50~(3821.84)$ for $f(R)$~($\Lambda$CDM) with $\chi^2_{f(R)} \lesssim \chi^2_{\Lambda CDM}$, indicating that exponential $f(R)$ is consistent with $\Lambda$CDM.
 The  neutrino mass sum is evaluated  to be $\Sigma m_\nu<0.06816~(0.06121)$ for $f(R)$~($\Lambda$CDM) at 68$\%$ C.L., which is relaxed at 11 $\%$ comparing with $\Lambda$CDM. This phenomenon is caused by the shortened age of the universe in the exponential $f(R)$ model, in which  the matter density fluctuation is suppressed as discussed in~\cite{Chen:2019uci} and Sec.~\ref{subsec:evol}.

To compare  exponential $f(R)$ gravity with $\Lambda$CDM, we introduce the Akaike Information Criterion (AIC)~\cite{Akaike:1974}, 
 Bayesian Information Criterion (BIC)~\cite{Schwarz:1978tpv}, and Deviance Information Criterion (DIC)~\cite{Spiegelhalter:2002yvw}.
The AIC is defined through the maximum likelihood $\mathcal{L}_{max}$ (satisfying $ -2 \, {\rm ln} \, {\mathcal L}_{max} \propto \chi ^2_{min}$ under 
the Gaussian likelihood assumption) and the number of model parameters, $d$:
\begin{align}
AIC = -2\, {\rm ln} \, {\mathcal L}_{max} + 2d = \chi ^2_{min} +2d.
\end{align}
The BIC is defined as
\begin{align}
BIC = -2\, {\rm ln} \, {\mathcal L}_{max} + d{\rm ln} N = \chi ^2_{min} + d {\rm ln} N,
\end{align}
where $N$ is the number of data points.

The DIC is determined by the quantities obtained from posterior distributions, given by
\begin{align}
DIC = D(\bar{\theta}) + 2 \,p_D,
\end{align}
where $D(\theta) = -2 \, {\rm ln} \, {\mathcal L}(\theta) + C, p_D = \overline{D(\theta)} - D(\overline{\theta}), C $ is a constant, and $p_D$ represents the effective number of parameters in the model.

We now compute the AIC, BIC, and DIC values from CMB+BAO+SN samples described above for both models, with the difference given by $\Delta AIC= AIC_{f(R)}-AIC_{\Lambda CDM}=1.66$, $\Delta BIC= BIC_{f(R)}-BIC_{\Lambda CDM}=7.85$, and $\Delta DIC= DIC_{f(R)}-DIC_{\Lambda CDM}=1.49$,
respectively.
The results are summarized in Table \ref{tab:AIC}, where the differences  are residuals with respect to the $\Lambda$CDM model.
Since $\Delta AIC,\,\Delta DIC) <2$, being small,
there is no strong preference between the exponential $f(R)$ and $\Lambda$CDM  models in terms of AIC and DIC~\cite{Rezaei:2021qpq}. 
However, 
as $6< \Delta BIC <10$, there is \emph{a strong evidence} against the exponential $f(R)$ model~\cite{Liddle:2007fy}. 

\section{Conclusions}

We have considered the exponential $f(R)$ gravity model without the spatial flatness assumption. We have derived the energy density ($\rho_{de}$) and pressure ($p_{de}$) of dark energy, and simplified the modified Friedmann equations into a second
order differential equation in Eqs.~(\ref{eq:ddyh})-(\ref{eq:J3}) with the involvement of the spatial curvature $K$.
In our numerical calculations, by modifying the {\bf{CAMB}} program for the exponential $f(R)$ model in open, flat and closed universe, we have studied  
the cosmological evolutions of the dark energy density parameter and equation of state. 
We have found that exponential $f(R)$ has a shortened age of the universe comparing with $\Lambda$CDM. 
To constrain the cosmological parameters of the exponential $f(R)$ model, we have used the {\bf{CosmoMC}} package to explore the parameter space. In particular, we have obtained  that  $\lambda^{-1}=0.42927^{+0.39921}_{-0.32927}$ at 68$\%$ C.L and $\Omega_K=-0.00050^{+0.00420}_{-0.00414}$ at 95$\%$ C.L. In addition, we have gotten that  
 $\chi^2=3821.50~(3821.84)$ for $f(R)$~($\Lambda$CDM) with $\chi^2_{f(R)} \lesssim \chi^2_{\Lambda CDM}$, which matches the previous work in the flat universe~\cite{Yang:2010xq}.

We have also evaluated  the AIC, BIC, and DIC values for the exponential $f(R)$ and  $\Lambda$CDM models. 
We have shown that the $\Lambda$CDM model is slightly more preferable in terms of BIC, but such  a preference has not been found based on the  AIC and DIC results.
We note that all values of AIC, BIC and DIC in  $f(R)$ are  smaller than those in $\Lambda$CDM in our study, whereas it is not the case in some models discussed 
in the literature, such as  those in Refs.~\cite{Zheng:2021uee, Rezaei:2021qpq, Liddle:2007fy}. 
It is known that different combinations of the data  may  draw to different conclusions due to the dimensional inconsistency~\cite{Liddle:2007fy}. 
Clearly, to compare models in cosmology, it is necessary to explore various probes along with different  data sets.

\section*{Acknowledgments}
This work was supported in part by 
the National Key Research and Development Program of China (Grant No. 2020YFC2201501).


\end{document}